\documentclass{article}

\usepackage{arxiv}

\usepackage{bm}
\usepackage{subcaption}
\usepackage{amsmath}
\usepackage{amsfonts}
\usepackage{listings}

\usepackage[svgnames]{xcolor}

\lstdefinelanguage{XML}
{
  basicstyle=\ttfamily\footnotesize,
  morestring=[b]",
  moredelim=[s][\bfseries\color{Maroon}]{<}{\ },
  moredelim=[s][\bfseries\color{Maroon}]{</}{>},
  moredelim=[l][\bfseries\color{Maroon}]{/>},
  moredelim=[l][\bfseries\color{Maroon}]{>},
  morecomment=[s]{<?}{?>},
  morecomment=[s]{<!--}{-->},
  commentstyle=\color{DarkOliveGreen},
  stringstyle=\color{blue},
  identifierstyle=\color{red}
}

\usepackage{color}

\usepackage{pgfplots}
\usepackage{pgfplotstable}
\pgfplotsset{
compat=newest, 
tick label style={font=\footnotesize}, 
}

\usepackage{overpic}

\usepackage{tikz}
\usetikzlibrary{mindmap,shadows}
\usetikzlibrary{arrows,intersections}
\usepackage{smartdiagram}

\usepackage{booktabs}
\usepackage[]{units}

\title{openCFS-Data: Implementation of the Stochastic Noise Generation and Radiation Model (SNGR)}

\author{ Stefan Schoder \\
	Group of Aeroacoustics and Vibroacoustics, IGTE\\
	TU Graz\\
	Inffeldgasse 18, 8010 Graz \\
	\texttt{stefan.schoder@tugraz.at} \\
}

\renewcommand{\shorttitle}{openCFS Software}

\usepackage{hyperref}
\hypersetup{
pdftitle={openCFS Software},
pdfsubject={preprint},
pdfauthor={Schoder, Roppert},
pdfkeywords={FEM Sotware, open Source},
}

\begin{document}
\maketitle

\begin{abstract}
	Preliminary aeroacoustic investigations in competitive industries require rapid numerical simulation techniques to gain initial insight into the flow and acoustic field. Although there are capabilities to resolve virtually all turbulence length scales, these techniques are often impractical in early stages of component development. Therefore, the flow field is typically assessed by a Reynolds-averaged Navier Stokes Simulation. Building upon the results of that flow simulation, a stochastic approach to reconstruct the turbulent velocity fluctuations. In conjunction with a hybrid aeroacoustic workflow, this approach is useful in early stage virtual prototyping of aeroacoustic applications. In this working paper, we present the SNGR algorithm of CFS-Data, the open-source pre-post-processing part of openCFS, with a focus on the computation of aeroacoustic sources.
\end{abstract}


%

\section{Introduction}
\label{sec:Intro} 
Within this contribution, we concentrate on the openCFS \cite{CFS} module \textit{openCFS-Data} \cite{CFSDAT}, the implemented SNGR method called \textit{syntheticTurbulence\_SNGR} and potential model applications. The method was initially applied in \cite{weitz2019approach,weitz2019numerical} and more details on a first application can be found there. Stochastic methods constitute a low-cost computational fluid dynamics (CFD) approach to reconstruct the turbulent velocity fluctuations using results Reynolds-averaged Navier-Stokes (RANS) simulations. This approach was introduced by B\'echara et al. \cite{bechara1994stochastic} in 1994 and is known as stochastic noise generation and radiation (SNGR). Regarding the model in \cite{weitz2019approach}, it was applied to a cavity noise simulation using a hybrid aeroacoustic workflow \cite{schoder2018aeroacoustic,schoder2019hybrid,schoder2020conservative}. The derivatives of the Lighthill's source term were computed by a radial basis function scheme \cite{schoder2020radial,schoder2020aeroacoustic}. Using recently developed equations, the method can potentially be applied to aeroacoustic formulations based on the acoustic potential \cite{schoder2022aeroacoustic,schoder2022cpcwe} in combination with a Helmholtz decomposition  \cite{schoder2020postprocessing,schoder2020postprocessing2,schoder2019helmholtz,schoder2022post}. Furthermore, this methodology can be valuable for automotive OEMs since it poses a fast estimate on the flow-acoustic properties of broadband noise excitation which can be applied as loading \cite{engelmann2020generic,freidhager2021simulationen,schoder2020numerical,weitz2019numerical,maurerlehner2022aeroacoustic}. Furthermore, the hybrid aeroacoustic workflow was found to be useful for fan noise computations \cite{schoder2020computational,tautz2018source,schoder2021application,tieghi2022machine,tieghi2023machine,schoder2022dataset}, the noise emissions of the turbocharger compressor \cite{freidhager2022applicability,kaltenbacher2020modelling,freidhager2020influences}, the acoustics of fluid-structure-acoustic-interaction processes \cite{schoder2020hybrid,valavsek2019application,zorner2016flow,schoder2021aeroacoustic,falk20213d,lasota2021impact,maurerlehner2021efficient,schoder2022learning,lasota2023anisotropic,schoder2022pcwe,kraxberger2022machine} and turbulent wave emissions of hydro-machinery based on the Francis99 data \cite{lenarcic2015numerical}. Potential nonphysical behavior generated with the source computation can be identified using \cite{schoder2022error}. For a literature overview on the developments connecting to SNGR, have a look at the article \cite{lafitte2014turbulence}.

\section{Stochastic Noise Generation and Radiation}
As proposed by Kraichnan \cite{kraichnan1970diffusion} and Karweit et al. \cite{karweit1991simulation} a spatially stochastic turbulent velocity field can be generated as a finite sum of $N$ statistically independent random Fourier mode
\begin{equation}
\begin{aligned}
& \boldsymbol{u}_{\mathrm{t}}(\boldsymbol{x})=2 \sum_{n=1}^{N} \tilde{u}_{n} \cos \left(\boldsymbol{k}_{n} \cdot \boldsymbol{x}+\psi_{n}\right) \boldsymbol{\sigma}_{n},
\end{aligned}
\label{eq:1}
\end{equation}
where $\boldsymbol{x}$ is the spatial position, and $\boldsymbol{k}_{n}, \tilde{u}_{n}, \psi_{n}$, and $\boldsymbol{\sigma}_{n}$ are the wave vector, the amplitude, the phase, and the direction of the $n^{\text {th }}$ mode, respectively. The wave vector $\boldsymbol{k}_{n}$ is randomly chosen on a sphere of radius $k_{n}$ to ensure isotropy. Additionally, incompressibility of the turbulent flow field implies $\partial u_{t i} / \partial x_{i}=0$ and hence
\begin{equation}
\boldsymbol{k}_{n} \cdot \boldsymbol{\sigma}_{n}=\text { for } \quad n=1, \ldots, N
\label{eq:2}
\end{equation}
The turbulent kinetic energy $K$ is computed as the statistical mean of $1 / 2 u_{t i} u_{t i}$ and may thus be written with Eq. (\ref{eq:1}) as
\begin{equation}
K=\sum_{n=1}^{N} \tilde{u}_{n}^{2} .
\label{eq:3}
\end{equation}
A homogeneous isotropic turbulence is characterized by a three-dimensional energy spectrum $E(k)$ which allows to compute the turbulent kinetic energy by
\begin{equation}
\int_{0}^{\infty} E(k) \mathrm{d} k=K .
\label{eq:4}
\end{equation}
Moreover, the spectrum $E(k)$ allows computing the rate of dissipation of turbulence energy $\epsilon$ by
\begin{equation}
2 \nu \int_{0}^{\infty} k^{2} E(k) \mathrm{d} k=\epsilon,
\label{eq:5}
\end{equation}
where $\nu$ is the kinematic viscosity. Discretizing Eq. (\ref{eq:4}) and combining it with Eq. (\ref{eq:3}) yields
\begin{equation}
\tilde{u}_{n}=\sqrt{E\left(k_{n}\right) \Delta k_{n}} .
\label{eq:5-1}
\end{equation}
The spectrum $E(k)$ used to simulate the complete spectral range is a von Kármán-Pao spectrum \cite{von1948progress,pao1965structure}
\begin{equation}
E(k)=\alpha \frac{u^{\prime 2}}{k_{\mathrm{e}}} \frac{\left(k / k_{\mathrm{e}}\right)^{4}}{\left[1+\left(k / k_{\mathrm{e}}\right)^{2}\right]^{17 / 6}} \exp \left[-2\left(\frac{k}{k_{\eta}}\right)^{2}\right],
\label{eq:6}
\end{equation}
where $k_{\eta}=\left(\epsilon / \nu^{3}\right)^{1 / 4}$ is the Kolmogorov wave number, and $u^{\prime}=\sqrt{2 K / 3}$ is the root-mean-square value of the velocity fluctuations. Assuming that the turbulent kinetic energy $K$ and the rate of dissipation of turbulence energy $\epsilon$ are known from the computational fluid dynamics (CFD) solution, and inserting Eq. (\ref{eq:7}) into Eq. (\ref{eq:4}) yields
\begin{equation}
\alpha=\frac{55}{9 \sqrt{\pi}} \frac{\Gamma\left(\frac{5}{6}\right)}{\Gamma\left(\frac{1}{3}\right)} \approx 1.45276 .
\label{eq:7}
\end{equation}
To compute the wave number $k_{\mathrm{e}}$ of the most energetic eddies corresponding to the maximum of the energy spectrum $E(k)$, Eq. (\ref{eq:7}) is inserted into Eq. (\ref{eq:5}), which leads to
\begin{equation}
\frac{u^{\prime 3}}{\epsilon}=\frac{2}{k_{\mathrm{e}}}\left[\alpha \Gamma\left(\frac{2}{3}\right)\right]^{-\frac{3}{2}} \approx 0.725 \frac{1}{k_{\mathrm{e}}}
\label{eq:8}
\end{equation}
when $k_{\mathrm{e}} / k_{\eta} \rightarrow 0$.
To achieve a better discretization of the power in the lower wave number range corresponding to the larger energy-containing eddies, a logarithmic distribution of the $N$ wave numbers is used. A logarithmic step of such a distribution reads
\begin{equation}
\Delta k_{1}=\frac{\log k_{N}-\log k_{1}}{N-1}
\label{eq:9}
\end{equation}
and hence the $n^{\text {th }}$ wave number $k_{n}$ is given by
\begin{equation}
k_{n}=\exp \left[\log k_{1}+(n-1) \log \Delta k_{\mathrm{l}}\right] .
\label{eq:10}
\end{equation}
Since this method only models spatial correlation but not temporal correlation, Bailly et al. \cite{bailly1999stochastic} included time-dependence in Eq. (\ref{eq:1}) to arrive at reasonable statistical properties. Hence, the turbulent velocity field is now computed as a sum of unsteady random Fourier modes
\begin{equation}
\boldsymbol{u}_{\mathrm{t}}(\boldsymbol{x}, t)=2 \sum_{n=1}^{N} \tilde{u}_{n} \cos \left(\boldsymbol{k}_{n} \cdot\left(\boldsymbol{x}-t \boldsymbol{u}_{\mathrm{c}}\right)+\psi_{n}+\omega_{n} t\right) \boldsymbol{\sigma}_{n},
\label{eq:11}
\end{equation}
where $\boldsymbol{u}_{\mathrm{c}}$ is the convection velocity and $\omega_{n}$ is the angular frequency of the $n^{\text {th }}$ mode. As opposed to the convection velocity $\boldsymbol{u}_{\mathrm{c}}$, which is a function of the known local mean flow, the angular frequency $\omega_{n}$ is a random variable. Kraichnan \cite{kraichnan1970diffusion} treated $\omega_{n}$ and $k_{n}$ as independent variables; however, aerodynamic noise generation cannot be treated satisfactorily with this supposition. Therefore, Bailly et al. \cite{bailly1999stochastic} proposed to draw $\omega_{n}$ from a distribution associated to a Gaussian probability density function
\begin{equation}
p_{n}(\omega)=\frac{1}{\omega_{0, n} \sqrt{2 \pi}} \exp \left(-\frac{\left(\omega-\omega_{0, n}\right)^{2}}{2 \omega_{0, n}^{2}}\right),
\label{eq:12}
\end{equation}
where the mean angular frequency of the $n^{\text {th }}$ mode $\omega_{0, n}$ is connected to the wave number $k$ by $\omega_{0, n}=u^{\prime} k_{n}$. Additionally, the authors propose to use the Heisenberg time $\tau_{\mathrm{H}} \sim\left(u^{\prime} k\right)^{-1}$ as the spectrally local characteristic time.
Billson et al. \cite{billson2004modeling} compute the wave number $k_{\mathrm{e}}$ of the most energetic eddies starting from the turbulence length scale of the RANS solution
\begin{equation}
\Lambda=C_{\mu} \frac{K^{3 / 2}}{\epsilon},
\label{eq:13}
\end{equation}
where $C_{\mu}$ is a closure coefficient of the $k-\epsilon$ turbulence model. Under the assumption that this length scale is the equal to the integral length scale for isotropic turbulence \cite{lafitte2014turbulence}, the following relation holds
\begin{equation}
\Lambda=\frac{\pi}{2 u^{\prime 2}} \int_{0}^{\infty} \frac{E(k)}{k} \mathrm{~d} k .
\label{eq:14}
\end{equation}
Hence, $k_{\mathrm{e}}$ can be computed as
\begin{equation}
k_{\mathrm{e}}=\frac{9 \pi}{55} \frac{\alpha}{\Lambda} .
\label{eq:15}
\end{equation}

\subsection{Filter definition}

The following XML-snippet illustrates a typical setting of the SNGR filter.

\begin{lstlisting}[language=XML]
    <syntheticTurbulence_SNGR id="synthTurb" tkeCriterion="0.1" 
    numOfModes="30" inputFilterIds="cfdEnsight" lengthScaleFactor="1.0" 
    timeScaleFactor="1.0" angularFreqFactor="1.0">
       <incrementModes>equidistant</incrementModes>
       <waveNumBounds minWN="0.1" maxWN="10"/>
       <frequencyBounds minFreq="200" maxFreq="12000"/>
       <TKE resultName="fluidMechTKE"/>
       <TEF resultName="fluidMechTEF"/>
       <localDensity resultName="fluiMechDensity"/>
       <localTemp resultName="fluidMechTemp"/>
       <meanVelocity resultName="fluidMechVelocity"/>
       <output resultName="syntheticTurbVelocity"/>
    </syntheticTurbulence_SNGR>
\end{lstlisting} 
\begin{itemize}
\item tkeCriterion: only use turbulent kinetic energy (TKE) values when larger than 10\%.
\item numOfModes: $N$
\item lengthScaleFactor, timeScaleFactor, angularFreqFactor scalars to stretch the space, time, or frequency respectively
\item incrementModes: equidistant or logarithmic
\item waveNumBounds: minimum and maximum resolve wave number
\item frequencyBounds: minimum and maximum resolve frequency
\item TKE, TEF (turbulent eddy frequency), localDensity, localTemp (field Temperature), meanVelocity are the field variables that must be supplied by the input.
\item output: output field definition.
\end{itemize}

\section{Acknowledge}
We would like to acknowledge the authors of openCFS and the implementation of M. Weitz.

\bibliographystyle{abbrv}
\bibliography{references}  

\begin{thebibliography}{10}

\bibitem{bailly1999stochastic}
C.~Bailly and D.~Juve.
\newblock A stochastic approach to compute subsonic noise using linearized
  euler's equations.
\newblock In {\em 5th AIAA/CEAS aeroacoustics conference and exhibit}, page
  1872, 1999.

\bibitem{bechara1994stochastic}
W.~Bechara, C.~Bailly, P.~Lafon, and S.~M. Candel.
\newblock {Stochastic approach to noise modeling for free turbulent flows}.
\newblock {\em AIAA journal}, 32(3):455--463, 1994.

\bibitem{billson2004modeling}
M.~Billson, L.-E. Eriksson, L.~Davidson, and P.~Jordan.
\newblock Modeling of synthesized anisotropic turbulence and its sound
  emission.
\newblock In {\em 10th AIAA/CEAS aeroacoustics conference}, page 2857, 2004.

\bibitem{engelmann2020generic}
R.~Engelmann, C.~Gabriel, S.~Schoder, and M.~Kaltenbacher.
\newblock A generic testbody for low-frequency aeroacoustic buffeting.
\newblock Technical report, SAE Technical Paper, 2020.

\bibitem{falk20213d}
S.~Falk, S.~Kniesburges, S.~Schoder, B.~Jakuba{\ss}, P.~Maurerlehner,
  M.~Echternach, M.~Kaltenbacher, and M.~D{\"o}llinger.
\newblock 3d-fv-fe aeroacoustic larynx model for investigation of functional
  based voice disorders.
\newblock {\em Frontiers in physiology}, 12:616985, 2021.

\bibitem{freidhager2021simulationen}
C.~Freidhager, P.~Maurerlehner, K.~Roppert, A.~Wurzinger, A.~Hauser,
  M.~Heinisch, S.~Schoder, and M.~Kaltenbacher.
\newblock Simulationen von str{\"o}mungsakustik in rotierenden bauteilen zur
  entwicklung von antriebskonzepten der autos der zukunft.
\newblock {\em e \& i Elektrotechnik und Informationstechnik}, 138(3):212--218,
  2021.

\bibitem{freidhager2020influences}
C.~Freidhager, S.~Schoder, and M.~Kaltenbacher.
\newblock The influences of spatial and temporal discretization in flow
  simulation on lighthill’s aeroacoustic source terms applied to a
  turbocharger.
\newblock In {\em AIAA AVIATION 2020 FORUM}, page 2546, 2020.

\bibitem{freidhager2022applicability}
C.~Freidhager, S.~Schoder, P.~Maurerlehner, A.~Renz, S.~Becker, and
  M.~Kaltenbacher.
\newblock Applicability of two hybrid sound prediction methods for assessing
  in-duct sound absorbers of turbocharger compressors.
\newblock {\em Acta Acustica}, 6:37, 2022.

\bibitem{kaltenbacher2020modelling}
M.~Kaltenbacher, C.~Freidhager, and S.~Schoder.
\newblock Modelling and numerical simulation of the noise generated by
  automotive turbocharger compressor.
\newblock Technical report, SAE Technical Paper, 2020.

\bibitem{karweit1991simulation}
M.~Karweit, P.~Blanc-Benon, D.~Juv{\'e}, and G.~Comte-Bellot.
\newblock Simulation of the propagation of an acoustic wave through a turbulent
  velocity field: A study of phase variance.
\newblock {\em The Journal of the Acoustical Society of America}, 89(1):52--62,
  1991.

\bibitem{kraichnan1970diffusion}
R.~H. Kraichnan.
\newblock Diffusion by a random velocity field.
\newblock {\em The physics of fluids}, 13(1):22--31, 1970.

\bibitem{kraxberger2022machine}
F.~Kraxberger, A.~Wurzinger, and S.~Schoder.
\newblock Machine-learning applied to classify flow-induced sound parameters
  from simulated human voice.
\newblock {\em arXiv preprint arXiv:2207.09265}, 2022.

\bibitem{lafitte2014turbulence}
A.~Lafitte, T.~L. Garrec, C.~Bailly, and E.~Laurendeau.
\newblock Turbulence generation from a sweeping-based stochastic model.
\newblock {\em AIAA journal}, 52(2):281--292, 2014.

\bibitem{lasota2021impact}
M.~Lasota, P.~{\v{S}}idlof, M.~Kaltenbacher, and S.~Schoder.
\newblock Impact of the sub-grid scale turbulence model in aeroacoustic
  simulation of human voice.
\newblock {\em Applied Sciences}, 11(4):1970, 2021.

\bibitem{lasota2023anisotropic}
M.~Lasota, P.~{\v{S}}idlof, P.~Maurerlehner, M.~Kaltenbacher, and S.~Schoder.
\newblock Anisotropic minimum dissipation subgrid-scale model in hybrid
  aeroacoustic simulations of human phonation.
\newblock {\em arXiv preprint arXiv:2301.00606}, 2023.

\bibitem{lenarcic2015numerical}
M.~Lenarcic, M.~Eichhorn, S.~Schoder, and C.~Bauer.
\newblock Numerical investigation of a high head francis turbine under steady
  operating conditions using foam-extend.
\newblock In {\em Journal of Physics: Conference Series}, volume 579, page
  012008. IOP Publishing, 2015.

\bibitem{maurerlehner2021efficient}
P.~Maurerlehner, S.~Schoder, C.~Freidhager, A.~Wurzinger, A.~Hauser,
  F.~Kraxberger, S.~Falk, S.~Kniesburges, M.~Echternach, M.~D{\"o}llinger,
  et~al.
\newblock Efficient numerical simulation of the human voice.
\newblock {\em e \& i Elektrotechnik und Informationstechnik}, 138(3):219--228,
  2021.

\bibitem{maurerlehner2022aeroacoustic}
P.~Maurerlehner, S.~Schoder, J.~Tieber, C.~Freidhager, H.~Steiner, G.~Brenn,
  K.-H. Sch{\"a}fer, A.~Ennemoser, and M.~Kaltenbacher.
\newblock Aeroacoustic formulations for confined flows based on incompressible
  flow data.
\newblock {\em Acta Acustica}, 6:45, 2022.

\bibitem{pao1965structure}
Y.-H. Pao.
\newblock Structure of turbulent velocity and scalar fields at large
  wavenumbers.
\newblock {\em The Physics of Fluids}, 8(6):1063--1075, 1965.

\bibitem{schoder2022cpcwe}
S.~Schoder.
\newblock cpcwe--perturbed convective wave equation based on compressible
  flows.
\newblock {\em arXiv preprint arXiv:2209.11410}, 2022.

\bibitem{schoder2022pcwe}
S.~Schoder.
\newblock Pcwe for fsai--derivation of scalar wave equations for
  fluid-structure-acoustics interaction of low mach number flows.
\newblock {\em arXiv preprint arXiv:2211.07490}, 2022.

\bibitem{schoder2022dataset}
S.~Schoder and F.~Czwielong.
\newblock Dataset fan-01: Revisiting the eaa benchmark for a low-pressure axial
  fan.
\newblock {\em arXiv preprint arXiv:2211.12014}, 2022.

\bibitem{schoder2020computational}
S.~Schoder, C.~Junger, and M.~Kaltenbacher.
\newblock Computational aeroacoustics of the eaa benchmark case of an axial
  fan.
\newblock {\em Acta Acustica}, 4(5):22, 2020.

\bibitem{schoder2020radial}
S.~Schoder, C.~Junger, K.~Roppert, and M.~Kaltenbacher.
\newblock Radial basis function interpolation for computational aeroacoustics.
\newblock In {\em AIAA AVIATION 2020 FORUM}, page 2511, 2020.

\bibitem{schoder2019hybrid}
S.~Schoder and M.~Kaltenbacher.
\newblock Hybrid aeroacoustic computations: State of art and new achievements.
\newblock {\em Journal of Theoretical and Computational Acoustics},
  27(04):1950020, 2019.

\bibitem{schoder2019helmholtz}
S.~Schoder, M.~Kaltenbacher, and K.~Roppert.
\newblock Helmholtz's decomposition applied to aeroacoustics.
\newblock In {\em 25th AIAA/CEAS Aeroacoustics Conference}, 2019-2561.

\bibitem{schoder2022aeroacoustic}
S.~Schoder, M.~Kaltenbacher, {\'E}.~Spieser, H.~Vincent, C.~Bogey, and
  C.~Bailly.
\newblock Aeroacoustic wave equation based on pierce's operator applied to the
  sound generated by a mixing layer.
\newblock In {\em 28th AIAA/CEAS Aeroacoustics 2022 Conference}, page 2896,
  2022.

\bibitem{schoder2022error}
S.~Schoder, F.~Kraxberger, S.~Falk, A.~Wurzinger, K.~Roppert, S.~Kniesburges,
  M.~D{\"o}llinger, and M.~Kaltenbacher.
\newblock Error detection and filtering of incompressible flow simulations for
  aeroacoustic predictions of human voice.
\newblock {\em The Journal of the Acoustical Society of America},
  152(3):1425--1436, 2022.

\bibitem{schoder2020numerical}
S.~Schoder, I.~Lazarov, and M.~Kaltenbacher.
\newblock Numerical investigation of a deep cavity with an overhanging lip
  considering aeroacoustic feedback mechanism.
\newblock {\em arXiv preprint arXiv:2006.03279}, 2020.

\bibitem{schoder2021aeroacoustic}
S.~Schoder, P.~Maurerlehner, A.~Wurzinger, A.~Hauser, S.~Falk, S.~Kniesburges,
  M.~D{\"o}llinger, and M.~Kaltenbacher.
\newblock Aeroacoustic sound source characterization of the human voice
  production-perturbed convective wave equation.
\newblock {\em Applied Sciences}, 11(6):2614, 2021.

\bibitem{schoder2022post}
S.~Schoder, E.~Museljic, F.~Kraxberger, and A.~Wurzinger.
\newblock Post-processing subsonic flows using physics-informed neural
  networks.
\newblock In {\em 2023 AIAA AVIATION Forum}, 2022.

\bibitem{schoder2022learning}
S.~Schoder and K.~Roppert.
\newblock Learning expertise actively to model domain knowledge (lead) with
  application to human phonation.
\newblock {\em arXiv}, 2022.

\bibitem{CFS}
S.~Schoder and K.~Roppert.
\newblock opencfs: Open source finite element software for coupled field
  simulation--part acoustics.
\newblock {\em arXiv preprint arXiv:2207.04443}, 2022.

\bibitem{CFSDAT}
S.~Schoder and K.~Roppert.
\newblock opencfs-data: Data pre-post-processing tool for
  opencfs--aeroacoustics source filters.
\newblock {\em arXiv preprint arXiv:2302.03637}, 2023.

\bibitem{schoder2020postprocessing2}
S.~Schoder, K.~Roppert, and M.~Kaltenbacher.
\newblock Helmholtz’s decomposition for compressible flows and its
  application to computational aeroacoustics.
\newblock {\em SN Partial Differ. Equ. Appl.}, pages 1--20, 2020.

\bibitem{schoder2020postprocessing}
S.~Schoder, K.~Roppert, and M.~Kaltenbacher.
\newblock Postprocessing of direct aeroacoustic simulations using helmholtz
  decomposition.
\newblock {\em AIAA Journal}, pages 1--9, 2020.

\bibitem{schoder2020aeroacoustic}
S.~Schoder, K.~Roppert, M.~Weitz, C.~Junger, and M.~Kaltenbacher.
\newblock Aeroacoustic source term computation based on radial basis functions.
\newblock {\em International Journal for Numerical Methods in Engineering},
  121(9):2051--2067, 2020.

\bibitem{schoder2020hybrid}
S.~Schoder, M.~Weitz, P.~Maurerlehner, A.~Hauser, S.~Falk, S.~Kniesburges,
  M.~D{\"o}llinger, and M.~Kaltenbacher.
\newblock Hybrid aeroacoustic approach for the efficient numerical simulation
  of human phonation.
\newblock {\em The Journal of the Acoustical Society of America},
  147(2):1179--1194, 2020.

\bibitem{schoder2021application}
S.~Schoder, A.~Wurzinger, C.~Junger, M.~Weitz, C.~Freidhager, K.~Roppert, and
  M.~Kaltenbacher.
\newblock Application limits of conservative source interpolation methods using
  a low mach number hybrid aeroacoustic workflow.
\newblock {\em Journal of Theoretical and Computational Acoustics},
  29(01):2050032, 2021.

\bibitem{schoder2018aeroacoustic}
S.~J. Schoder.
\newblock {\em Aeroacoustic analogies based on compressible flow data}.
\newblock PhD thesis, Wien, 2018.

\bibitem{schoder2020conservative}
S.~J. Schoder, C.~Junger, M.~Weitz, and M.~Kaltenbacher.
\newblock Conservative interpolation of aeroacoustic sources in a hybrid
  workflow applied to fan.
\newblock {\em arXiv preprint arXiv:2009.02341}, 2020.

\bibitem{tautz2018source}
M.~Tautz, K.~Besserer, S.~Becker, and M.~Kaltenbacher.
\newblock Source formulations and boundary treatments for lighthill’s analogy
  applied to incompressible flows.
\newblock {\em AIAA Journal}, 56(7):2769--2781, 2018.

\bibitem{tieghi2022machine}
L.~Tieghi, S.~Becker, A.~Corsini, G.~Delibra, S.~Schoder, and F.~Czwielong.
\newblock Machine-learning clustering methods applied to detection of noise
  sources in low-speed axial fan.
\newblock In {\em 2022 Turbomachinery Technical Conference \& Exposition: ASME
  Turbo Expo 2022}, 2022.

\bibitem{tieghi2023machine}
L.~Tieghi, S.~Becker, A.~Corsini, G.~Delibra, S.~Schoder, and F.~Czwielong.
\newblock Machine-learning clustering methods applied to detection of noise
  sources in low-speed axial fan.
\newblock {\em Journal of Engineering for Gas Turbines and Power},
  145(3):031020, 2023.

\bibitem{valavsek2019application}
J.~Val{\'a}{\v{s}}ek, M.~Kaltenbacher, and P.~Sv{\'a}{\v{c}}ek.
\newblock On the application of acoustic analogies in the numerical simulation
  of human phonation process.
\newblock {\em Flow, Turbulence and Combustion}, 102(1):129--143, 2019.

\bibitem{von1948progress}
T.~Von~Karman.
\newblock Progress in the statistical theory of turbulence.
\newblock {\em Proceedings of the National Academy of Sciences},
  34(11):530--539, 1948.

\bibitem{weitz2019approach}
M.~Weitz.
\newblock {\em An approach to compute cavity noise using stochastic noise
  generation and radiation}.
\newblock PhD thesis, Wien, 2019.

\bibitem{weitz2019numerical}
M.~Weitz, S.~Schoder, and M.~Kaltenbacher.
\newblock Numerical investigation of the resonance behavior of flow-excited
  helmholtz resonators.
\newblock {\em PAMM}, 19(1):e201900033, 2019.

\bibitem{zorner2016flow}
S.~Z{\"o}rner, P.~{\v{S}}idlof, A.~H{\"u}ppe, and M.~Kaltenbacher.
\newblock Flow and acoustic effects in the larynx for varying geometries.
\newblock {\em Acta Acustica united with Acustica}, 102(2):257--267, 2016.

\end{thebibliography}






\end{document}